\documentclass[12pt, manuscript]{aastex}

\begin{document}

\title{Hard X-ray Emission Associated with White Dwarfs}

\author{Ian J.\ O'Dwyer, You-Hua Chu, Robert A.\ Gruendl, Mart\'{\i}n 
A.\ Guerrero, Ronald F.\ Webbink}
\affil{Department of Astronomy, University of Illinois,
    1002 W. Green Street,
    Urbana, IL 61801; iodwyer@astro.uiuc.edu, chu@astro.uiuc.edu,
    gruendl@astro.uiuc.edu, mar@astro.uiuc.edu, webbink@astro.uiuc.edu}

\begin{abstract}

Inspired by the hard X-ray emission from WD\,2226$-$210, the
central star of the Helix Nebula, we have made a systematic 
search for similar sources by correlating the white dwarf 
catalog of \citet{MCC99} and the $ROSAT$ PSPC point source 
catalog of \citet{WGA00}.  We find 76 white dwarfs coincident 
with X-ray sources at a high level of confidence.  Among these 
sources, 17 show significant hard X-ray emission at energies 
$>$ 0.5 keV.  Twelve of these white dwarfs with hard X-ray 
emission are in known binary systems, in two of which the 
accretion of the close companion's material onto the white 
dwarf produces hard X-ray emission, and in the other ten of 
which the late-type companions' coronal activity emits hard 
X-rays.  One apparently single white dwarf is projected near 
an AGN which is responsible for the hard X-ray emission.
The remaining four white dwarfs and two additional white dwarfs
with hard X-ray emission appear single.
The lack of near-IR excess from the apparently single white
dwarfs suggests that either X-ray observations are more
effective than near-IR photometry in diagnosing faint
companions or a different emission mechanism is needed.
It is intriguing that 50\% of the six apparently single 
white dwarfs with hard X-ray emission are among the hottest
white dwarfs.  
We have compared X-ray properties of 11 hot white dwarfs with
different spectral types, and conclude that stellar pulsation
and fast stellar winds are not likely the origin of the hard
X-ray emission, but a leakage of the high-energy Wien tail 
of emission from deep in the stellar atmosphere
remains a tantalizing source of hard X-ray emission from hot
DO and DQZO white dwarfs.
A complete survey using the entire $ROSAT$ PSPC archive is needed 
to enlarge the sample of white dwarfs with hard X-ray emission.
Follow-up near-IR photometric observations are needed to verify 
the existence of late-type companions and high-resolution deep 
X-ray observations are needed to verify the positional coincidence 
and to study the X-ray spectral properties in order to determine
the origin and nature of the hard X-ray emission.

\end{abstract}


\section{Introduction}

White dwarfs can be sources of soft ($<$ 0.4 keV) 
X-ray emission if their atmospheres have high temperatures and
low opacities: $T_{\rm eff} \gtrsim$ 23,000 K for DA white dwarfs with 
pure hydrogen atmospheres \citep{JOR94}, 23,000 K $\lesssim T_{\rm eff} 
\lesssim$ 54,000 K for DA white dwarfs containing significant quantities 
of heavy elements in their atmospheres \citep{MAR97}, and 
$T_{\rm eff} \gtrsim$ 100,000 K for DO and PG~1159 white dwarfs with 
helium-rich atmospheres \citep{MOT93}.
$ROSAT$ observations of such X-ray sources show soft spectra
rising toward $ROSAT$'s low-energy limit at 0.1 keV.
The DO white dwarf KPD~0005+5106 (= WD\,0005+511), with its 
spectrum peaking at 0.2 keV, appears to pose an exception and
has been interpreted as having cool ($2-3 \times 10^5$ K) coronal 
emission \citep{FLE93}.
{\it No hard X-ray ($>$ 0.5 keV) emission is expected from single 
white dwarfs.}

It is thus puzzling that WD\,2226$-$210\footnote{This white dwarf 
was cataloged with a sign error in declination as WD\,2226$+$210 
by \citet{MCC99}.}, the 103,600 K DAO white 
dwarf \citep{Metal88,Nap99} in the Helix Nebula, appears to be
single \citep{CIA99}, but has hard X-ray emission.
$ROSAT$ observations of WD\,2226$-$210 have detected not only a soft 
spectral component as expected from the white dwarf's photosphere, 
but also a hard spectral component peaking at 0.8 keV \citep{LEA94}.
Recent $Chandra$ observations have confirmed that the hard X-ray 
emission is unresolved and coincident with the white dwarf; 
however, the luminosity and variability of the hard X-ray 
emission is similar to that of a dMe star \citep{GUE01}.
Follow-up spectroscopic observations by \citet{GRU01} detected 
variability in the H$\alpha$ line profile of WD\,2226$-$210,
suggesting the presence of a companion.
It is thus possible that WD\,2226$-$210 has an X-ray-emitting dMe 
companion which is too faint and close to the white dwarf to be 
detected at visible wavelengths. 
Based on the $I$ magnitude of WD\,2226$-$210 reported by
\citet{CIA99} and the $J$, $H$, and $K$ magnitudes from
the 2MASS Survey (see Table 3 below), we estimate that the 
hypothetical companion must have a spectral type later than
M5-6\,V.
 
Hard X-ray emission from white dwarfs may be used to infer the 
presence of a binary companion.  Indeed, \citet{Fetal96} found 9 
DA white dwarfs with hard X-ray emission using the $ROSAT$ All-Sky 
Survey, and all of these 9 white dwarfs have late-type (F, G, K, 
and M) companions. 
This has motivated us to search for other white dwarfs 
which exhibit hard X-ray emission.
We have found that 94 white dwarfs cataloged by \citet{MCC99}
appear coincident with $ROSAT$ X-ray point sources in the WGA 
catalog \citep[hereafter WGACAT]{WGA00}.
To confirm the positional coincidence, we have downloaded the 
$ROSAT$ data, and compared these X-ray images with optical 
images of the white dwarfs.
We have further extracted the $ROSAT$ spectra to examine
whether hard X-ray emission is present.
For white dwarfs associated with hard X-ray emission, we have used
the literature and near-IR photometry to assess the existence of
binary companions, and further investigate the origin of
hard X-ray emission from the apparently single white dwarfs.
This paper reports the results of our study.
The sample of white dwarfs with hard X-ray emission and our method 
of analysis are described in Section 2, the binarity status of the
white dwarfs with hard X-ray emission is reported in Section 3, and 
the origin and implications of the hard X-ray emission associated 
with white dwarfs is discussed in Section 4.  A summary is given 
in Section 5.

\section{Search for Hard X-ray Sources Associated with White Dwarfs}

To search for X-ray emission from white dwarfs, we have
correlated the white dwarfs from the most recent catalog by 
\citet{MCC99} with the X-ray sources from the WGACAT, a point 
source catalog generated from all $ROSAT$ PSPC pointed 
observations made without the boron filter.
For an initial identification of positional coincidence, we 
require better than 1$'$ agreement between the cataloged 
positions of the white dwarf and X-ray point source.  
This is a conservative criterion because: (1) the on-axis point 
spread function (PSF) of the $ROSAT$ Position Sensitive Proportional 
Counter (PSPC) is $\sim$40$''$ at 1 keV and much worse near 0.1 keV 
\citep{SNO94,CHU93}, (2) the coordinates of the white dwarf might
be uncertain by up to 30$''$, and (3) the proper motions of the
white dwarfs have not been considered and in some cases may be 
large enough for the star to move more than 1$'$ between the 
epochs when the optical and X-ray observations were made.

We find 94 white dwarfs that appear to be associated with point 
X-ray sources. 
To confirm the positional coincidence and to examine further
the X-ray spectra of these sources, we have retrieved the
event files of their $ROSAT$ PSPC observations from the
High Energy Astrophysics Science Archive Research Center
(HEASARC) at NASA's Goddard Space Flight Center.
The PSPC event files are used to extract broad-band 
(0.1--2.4 keV) X-ray images.
These X-ray images are smoothed with a Gaussian with $\sigma$
= 15$''$ and the resulting X-ray
contours at 10, 20, 50, 70, and 90\% of the peak intensity 
are overplotted on the broad-band optical images retrieved
from the Digitized Sky Survey (DSS).  To identify the white
dwarfs, we have used the finding charts provided by J. Holberg
at http://procyon.lpl.arizona.edu/WD/.

The results of our detailed comparison of the positions of X-ray 
sources and white dwarfs are presented in Table 1.  Columns 
1--3 give the identifications, common names, and spectral types 
of the white dwarfs; columns 4--6 list the corresponding X-ray
sources in the WGACAT, the $ROSAT$ PSPC observations used for 
the detection, and the exposure times of the PSPC observations;
column 7 describes the positional coincidence between the 
X-ray source and the white dwarf; and columns 8--10 present
the counts detected in the soft (0.1--0.4 keV), medium
(0.4--0.9 keV), and hard (0.9--2.0 keV) bands, as reported in
the WGACAT.

Eighteen of the 94 coincidences we initially identified could 
not be confirmed, noted as U1--4 in column 7 of Table 1, because 
the X-ray sources are not convincingly centered on the white 
dwarf (U1), the positions of the X-ray sources are compromised 
by the occultation of the PSPC window support structure or the 
superposition of a bright background (U2), multiple candidates 
of optical counterparts are present within the PSPC PSF (U3),
or the X-ray sources are too faint to be credible (U4).

One notable example of a U4 mis-identification is WD\,1910+047,
which was discovered and suggested by \citet{MBA87} to be 
associated with an {\it Einstein} X-ray source.
The weak WGACAT source J1912.5+0452 located within 1$'$ 
from WD\,1910+047 was detected in a 2.8 ks PSPC observation
(rp400271n00) at a 44$'$ off-axis position, but not confirmed in 
a 20 ks PSPC observation (rp400271a01) with identical pointing.
We have examined this shallow PSPC observation and 
concluded that this source is spurious.  Using a deep PSPC 
observation (rp500058a02, 20.6 ks) centered at 19$'$ from 
WD\,1910+047, we find an X-ray source near the white dwarf, but 
it is coincident with a star $\sim$80$''$ SE of the white dwarf 
and the X-ray spectrum is typical for coronal emission from a 
late-type star (see Figure 1).  This result explains the 
conclusion of the model atmosphere analysis by \citet{Ve90} 
that the {\it Einstein} X-ray source is too luminous for the 
white dwarf.

There remain 76 coincidences that are confirmed with a high 
degree of confidence.
We have further extracted X-ray spectra from the PSPC event 
files for these X-ray sources coincident with white dwarfs.
We find 69 dominated by soft X-ray emission ($<$0.5 keV),
as expected from white dwarfs; however, 10 of these also 
exhibit hard X-ray emission (0.5--2.4 keV).
The spectra of the remaining 7 sources have characteristics 
consistent with coronal emission at temperatures of a few 
$\times10^6$ K.
These PSPC spectra appear ``double-peaked" because of the carbon 
K$\alpha$ absorption of the PSPC entrance window at 0.24--0.4 keV;
the hard component ($>$0.5 keV) is at least 25\% as strong as the 
soft component ($<$0.5 keV).

The 17 white dwarfs associated with hard X-ray emission are 
listed in Table 2, and their images and X-ray spectra are 
presented in Figure 2.  The left panels display DSS images 
centered on the white dwarfs and overlaid by X-ray contours 
to illustrate the positional coincidence of the white dwarf 
and the X-ray source.  The white dwarf position is marked 
when several stars are projected in its vicinity.
The right panels display the PSPC 
spectra of the X-ray sources.  For objects overwhelmed by 
soft X-ray emission, we also plot the spectra with an expanded 
Y-scale to show the hard component.  The PSPC observations 
listed in Table 1 have been used to extract these X-ray images 
and spectral information.

Note that Column 10 of Table 1 shows many more white dwarfs 
with 3$\sigma$ detections in the 0.9--2.4 keV band.  
However, all except the above 17 white dwarfs are rejected 
because their coincidences with X-ray source are not 
confirmed (U1-4 in Column 7) or the WGACAT source counts 
are contaminated/confused with background hard X-ray sources.
For example, some white dwarfs are observed at such large 
off-axis angles ($>40'$) that the PSF contains background 
sources (e.g., WD\,0037+312 and WD\,1040+451), and some are 
blended with adjacent sources (e.g., WD\,1821+643, 
WD\,1844$-$654, and WD\,0904+511).
Under such circumstances, the source detection algorithm of 
WGACAT does not effectively exclude contaminating background 
sources within the PSF, resulting in the apparent detection of
hard X-ray emission.

\section{Binarity of White Dwarfs with Hard X-ray Emission}

White dwarfs by themselves do not emit hard ($>$ 0.5 keV)
X-rays.
If a white dwarf accretes material from its surroundings, the
gravitational energy released may power hard X-ray emission.
In general, we can rule out the accretion of interstellar  
material because white dwarfs usually are not in a dense 
interstellar environment.  
We can also rule out the accretion of the planetary nebula 
produced by a white dwarf's progenitor, as it either has 
dissipated already or is expanding away at a speed much greater
than the escape velocity.  
Thus, the most likely source to provide material for accretion
onto a white dwarf is a binary companion.
Alternatively, if a white dwarf has a binary companion with
coronal activity, the hard X-ray emission from the companion
would appear to be associated with the white dwarf.
We therefore suspect that white dwarfs with hard X-ray emission
are in binary systems.

A literature search of the 17 white dwarfs with 
hard X-ray emission reveals that 12 of them are known 
binaries.  Below we describe these 12 known binary systems in 
\S 3.1, and the 5 apparently single white dwarfs in \S 3.2.

\subsection{Known Binaries}

{\it WD\,0216$-$032 (VZ Cet)} is a detached companion of Mira,
a pulsating, cool giant (M2-7\,III).  
\citet{Ketal97} measured a separation and position angle 
$\rho$ = 0\farcs578, $\theta$ = 108.3$^\circ$ (Ep.\ 1995.9424).
\citet{B80} estimated an orbital period of 400 yr for this
system (separation $a$ = 0\farcs85), but this result is 
extremely uncertain \citep{Metal01}.
The white dwarf flickers \citep{W72}, and $IUE$ observations 
suggest that it may possess an accretion disk \citep{RC85}.
The PSPC spectrum shows X-ray emission peaking at 1 keV.
This hard X-ray emission probably originates from the accretion,
since no single cool giants similar to Mira are known X-ray sources.

{\it WD\,0347+171 (V471~Tau)}  is in the 
eclipsing binary system V471 Tau with a period of 0.521 days
\citep{NY70,GS84}.  
The K2\,V companion is known for its coronal activity, and 
coronal mass ejections have been implied from the observations
of transient absorption features in the \ion{Si}{3} $\lambda$1206 
resonance line \citep{BMOS01}.  Using X-ray and EUV eclipses
of V471 Tau, \citet{Betal92} have demonstrated that the hard X-ray 
emission indeed originates from the K2\,V star.

{\it WD\,0429+176 (HZ~9)} is in
a spectroscopic binary system with a dM4.5e companion and
a period of 0.564 days \citep{LP81,GS84}.  Late-type
main sequence M stars, particularly the ones with Balmer lines
in emission (i.e., dMe stars), have been shown to be X-ray 
emitters \citep{Ru84}.  The dM4.5e companion of WD\,0429+176
is most likely responsible for the hard X-ray emission detected.

{\it WD\,0736+053 (Procyon B)} is a visual binary
companion of Procyon A, an F5\,IV-V star, with a period of 40.8 yr
\citep{Gi00}.  The coronal activity of Procyon A is responsible
for the hard X-ray emission \citep{Le89}.
In Figure 2, the wavy line shows the motion (proper and orbital) 
of the white dwarf between the  epochs of DSS and $ROSAT$ PSPC 
observations.

{\it WD\,1213+528 (EG~UMa)} was observed at a large 
off-axis angle.  The PSPC PSF extends over almost 2$'$, but 
the white dwarf appears to be the most likely optical counterpart
of the X-ray source.   WD 1213+528 has a dM2e binary companion 
in a 0.668 day period \citep{La82}.

{\it WD\,1255+258J (HD 112313B)} is the central star of the
planetary nebula LoTr\,5, and has a G5\,III companion with a
rotational period of 5.9 days \citep{Jetal96}. 
The orbital period of this binary is unknown, but probably in 
excess of one year; the binary is unresolved.
The strong \ion{Ca}{2} H \& K emission lines and broad variable
H$\alpha$ line of the G5\,III companion suggest coronal activity.
Thus the G5\,III companion is probably responsible for the observed
hard X-ray emission.

{\it WD\,1314+293 (HZ~43A)} has a dM3.5e companion \citep{NBF93}, 
resolved at $\rho$ = 2\farcs2, $\theta$ = 263$^\circ$ 
\citep[Ep.\ 1996;][]{Metal01}. 
This companion is likely responsible for the hard X-ray 
emission detected.

{\it WD\,1631+781 (1ES~1631+78.1)} has an unresolved dM4e 
companion \citep{Cetal95}.
From the lack of detectable radial velocity variations 
\citep{Setal95}, we conclude that the orbital period probably 
exceeds one year.  
Fluctuations in H$\beta$ emission from the dM4e companion
mark it as the probable site of hard X-ray emission.

{\it WD\,1633+572 (GJ~630.1B)} is the common proper motion
companion ($\rho$ = 26\farcs19, $\theta$ = 22.46$^\circ$
at Ep. 2000.0 from an astrometric fit to six DSS plates)
of the variable star CM~Dra \citep{Gr86}.
CM~Dra is itself an eclipsing binary containing two dM3-4e 
stars with a period of 1.26 days \citep{La77}.  
The proper motion of CM~Dra is so large that it must be 
considered when comparing the DSS and $ROSAT$ PSPC images.  
In Figure 2, we have drawn an arrow to show the proper motion 
of CM~Dra between the optical and X-ray epochs.
The X-ray source is centered on CM~Dra, rather than the
white dwarf; therefore, the hard X-ray emission clearly 
originates from the dM3-4e binary.

{\it WD\,1634$-$573 (HD~149499B)} is in a wide binary system
($\rho$ =  1\farcs319, $\theta$ = 51.71$^\circ$ at Ep.\ 1991.25,
from Hipparcos) with a K2\,Ve companion \citep{We79,We81}.  
The emission lines in the companion's spectrum are 
indicative of coronal activity, which may be responsible 
for the hard X-ray emission.

{\it WD\,1944$-$421 (V3885~Sgr)} is a cataclysmic variable
with orbital period 0.216 days \citep{Detal01}.
Among our sample of 17 white dwarfs with hard X-rays, this is 
the only one whose hard X-ray component is stronger than the 
soft X-ray component, indicating a different emission mechanism.
The X-ray emission from WD\,1944$-$421 must originate from the 
accretion of material from a binary companion onto the surface
of the white dwarf \citep{Pa94}.

{\it WD\,2154$-$512 (GJ~841B)} is a visual companion of GJ~841A
($\rho$ = 27\farcs93, $\theta$ = 251.5$^\circ$, Ep.\ 1987.23, 
from measurements on 2 DSS plates), which consists of two 
chromospherically active dM3--5e stars with an orbital period 
of 1.124 days \citep{JB93}.  
The dMe stars in GJ~841A are most likely responsible for
the hard X-ray emission detected.

\subsection{The Apparently Single White Dwarfs}

{\it WD\,0339$-$451} has an X-ray spectrum similar to those
associated with stellar coronal emission.  The temperature of 
this DA white dwarf is unknown, but its photospheric emission
probably does not contribute much to the X-ray emission detected
since the spectrum below 0.5 keV does not rise toward 0.1 keV
as expected.

{\it WD\,1134+300 (GJ~433.1)} is a DA2 white 
dwarf that has been used as a spectrophotometric standard star 
\citep{MG90, HLL01}.  Its PSPC spectrum shows a distinct
peak at 0.8--0.9 keV.  Detailed inspection shows that the soft 
X-ray emission is centered at the white dwarf, but the hard 
X-ray emission is centered at a position $\sim$41$''$ northwest
of the white dwarf, where an AGN has been identified by
\citet{Metal00}.  This background AGN must be responsible for 
the hard X-ray emission.

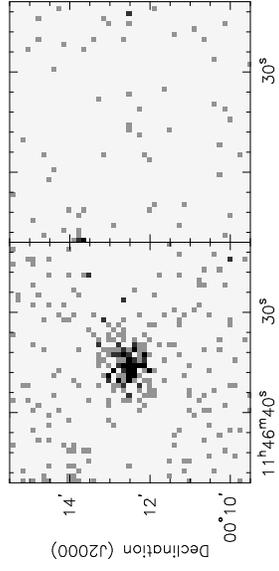
{\it WD\,1159$-$034 (GW~Vir)}, the prototype of PG\,1159 white
dwarfs, is a well-known pulsating variable \citep{WHV83}.  
The faint hard X-ray emission is centered on the white dwarf, 
but the spectrum is too noisy for detailed spectral analysis.

{\it WD\,1234+481 (PG~1234+481)} does not have X-ray emission
peaking at $\sim$0.9 keV as the other white dwarfs associated
with hard X-ray emission do.  However, the spectrum of 
WD\,1234+481 appears to show excess emission at 0.4--0.6 keV.

{\it WD\,1333+510 (PG~1333+510)} is detected at a large
off-axis angle, but it is the closest source, among three, to 
the peak of the X-ray emission.  The X-ray spectrum is similar
to those of stellar coronae.

\section{Discussion}

X-ray emission from white dwarfs is expected to be soft, but
roughly 20\% of X-ray sources associated with white dwarfs 
exhibit a hard X-ray component ($>$0.5 keV).
What is the origin of hard X-ray emission associated with
white dwarfs?
We initially speculated that all white dwarfs with hard 
X-ray emission possess binary companions, but our literature 
search reveals that 12 of the 17 white dwarfs with hard 
X-ray emission are known to be in binary or multiple systems,
one is superposed by chance near a background AGN, and the
remaining four are apparently single white dwarfs.
Below we argue that the binary companions are indeed directly
or indirectly responsible for the hard X-ray emission associated
with white dwarfs in binary systems, use 2MASS photometry to 
assess the existence of late-type companions, and discuss the 
hard X-ray emission from the hottest apparently single white 
dwarfs.  

\subsection{Hard X-ray Emission Associated with White Dwarfs 
in Binary Systems}

Two of our 12 white dwarfs in binary systems are known to accrete 
material from their companions: Mira and the cataclysmic variable
V3885 Sgr.  
The accretion of the companion's material onto the surface of 
the white dwarf produces hard X-ray emission \citep{Pa94}.
Of the ten non-accreting systems, six have one or two dMe 
companions, two have a K2\,V companion, one has an F5\,IV-V 
companion, and one has a G5\,III companion.
These late-type companions either are known for their active 
coronae or have emission lines indicating coronal activity;
therefore, these companions are most likely responsible for 
the observed hard X-ray emission.
In some cases, the origin of the hard X-ray emission has been
unambiguously established to be the late-type companions by
either positional coincidence (e.g., WD\,1633+572) or 
observations of eclipses (e.g., WD\,0347+171).

To confirm that these late-type companions are the source of 
the hard X-ray emission, we compare the hard X-ray 
luminosities and plasma temperatures of the binary systems 
with those expected from single late-type stars.
We have fitted thin plasma emission models \citep{RS77} to 
the 0.5--2.4 keV portion of the spectra for the objects 
with sufficient counts in this energy band.  
The resultant plasma temperature $kT$ and X-ray luminosity 
$L_{\rm X}$ in the 0.5--2.4 keV band are given in Table 2.
The $L_{\rm X}$ of white dwarfs with late-type companions
are completely consistent with those seen in dM and K stars, 
10$^{27}$--10$^{29}$ ergs~s$^{-1}$ \citep{SFG95,MMP00}.  
The best-fit plasma temperatures are also in the range 
for stellar coronae.  We further assess whether coronal 
activities of the late-type companions may be induced by 
binary interaction by correlating the projected binary
separation, $a$, with the X-ray luminosity
(see Table 2).  No clear correlation is seen.  This 
lack of correlation may be caused by the small number of
binary systems in our sample and the wide range of 
physical parameters involved.
The cataclysmic variable WD\,1944$-$421 (V3885 Sgr) has the 
highest hard X-ray luminosity, which is also in accord with 
the expectation of accretion from a Roche-lobe-filled 
companion.
Thus, the hard X-ray emission from white dwarfs in known binary
systems can be explained by the presence of their companions.

\subsection{Apparently Single White Dwarfs}

The remaining four white dwarfs associated with hard X-ray 
emission appear to be single.
If these white dwarfs contain previously-unknown, late-type 
companions, near-IR photometry may reveal these companions 
\citep[e.g.,][]{GAN00}.  
We have obtained $JHK$ photometric measurements of 
white dwarfs with hard X-rays available in the second incremental 
data release of the Two Micron All Sky Survey (2MASS), and listed 
them in Table 3, along with optical photometry from \citet{MCC99}.
In the case of WD\,1633+572 where the white dwarf and the binary 
dMe companions are resolved by 2MASS, separate entries are given.
To enlarge the sample for comparison, we have added two additional
white dwarfs with hard X-ray emission that are not from our survey:
WD\,0005+511 (= KPD~0005+5106; its hard X-ray emission will be
discussed in \S4.3) and WD\,2226$-$210 (= the central star of 
the Helix Nebula; its hard X-ray emission is described in \S1).

Clear near-IR excess is observed in the four known binary 
systems, but not in the apparently single white dwarfs.
The lack of near-IR excess, in conjunction with known distances,
places constraints on the possible companions of the white dwarfs.
The constraint is more stringent for white dwarfs at smaller 
distances because it would be harder to hide a companion.
The nearest apparently single white dwarf with hard X-ray emission
is WD\,1134+300 at 15.3 pc.
We estimate that its near-IR magnitudes can hide only a brown
dwarf companion more than 3 mag fainter than an M7\,V star.
Therefore, WD\,1134+300 has no stellar-mass companions, and the 
AGN projected within the PSPC PSF \citep{Metal00} is solely 
responsible for the hard X-ray emission observed.

Two other white dwarfs with hard X-ray emission have known
distances, WD\,1234+481 and WD\,2226$-$210.
Their lack of near-IR excess indicates that WD\,1234+481 can 
hide a companion of spectral type M7\,V or later, and
WD\,2226$-$210 later than M5-6\,V. 
If these two apparently single white dwarfs indeed have faint,
late-type companions, then the hard X-ray emission detected by 
$ROSAT$ is more effective than the near-IR excess detected by
2MASS in diagnosing the existence of a faint late-type companion.

\subsection{Hard X-ray Emission from the Hottest Apparently 
  Single White Dwarfs}

Two of the apparently single white dwarfs associated with hard 
X-rays from our survey are among the hottest known: 
WD\,1159$-$034 and WD\,2226$-$210. 
To compare these two hot white dwarfs to the one that has been
suggested to possess a corona, KPD~0005+5106, we have retrieved 
an archival $ROSAT$ PSPC pointed observation (rf200428n00) that 
was made with the boron filter for an exposure time of 5 ks.
As shown in Figure 3, the PSPC spectrum of KPD\,0005+5106 shows
not only the soft atmospheric emission below 0.5 keV, but also
hard X-ray emission near 1 keV.
This hard X-ray emission was not detected previously by
\citet{FLE93} using the $ROSAT$ All-Sky Survey 
observation of exposure time 504 s.
As the 5 ks pointed observation detected only 24$\pm$6
counts in the 0.5--2.4 keV band, if the hard X-ray flux is
constant, we expect only 2.4$\pm$1.5 hard X-ray counts 
in a 504 s exposure.
The non-detection of the hard component by \citet{FLE93}
thus does not imply a temporal variation.
The presence of hard X-ray emission from KPD\,0005+5106 
requires a plasma temperature at least 10$^6$ K, much higher 
than that suggested by \citet{FLE93}, and may be responsible 
for photoionizing \ion{O}{8} and producing the recombination 
lines observed \citep{WH92,SD92}. 
We have examined the 2MASS $JHK$ photometric data for 
KPD~0005+5106 (see Table 3), and find no evidence for a near-IR
excess.  Thus, KPD~0005+5106 is another apparently single hot 
white dwarf with hard X-ray emission.

While our statistical sample is extremely limited, it is
intriguing that $\sim$50\% of the apparently single white dwarfs 
with hard X-ray emission are among the hottest known white dwarfs:
120,000 K for WD\,0005+511 \citep[=KPD~0005+5106;][]{WHF94}, 
140,000 K for WD\,1159$-$034 \citep[= PG~1159$-$034;][]{DH98}, and 
103,600 K for WD\,2226$-$210 \citep[= CSPN of the Helix;][]{Nap99}.
Is it possible that the hard X-ray emission is the high-energy
Wien tail of the blackbody emission from deep in the stellar
atmosphere?  
It is beyond the scope of this paper to model the atmospheres of 
these white dwarfs and answer this question theoretically.
Instead, we will examine X-ray observations of the hottest white 
dwarfs to search for trends in their hard X-ray properties in 
order to gain insight into the origin of their hard X-ray emission.

We first examine X-ray properties of PG~1159 stars.  
Many pointed $ROSAT$ PSPC observations of PG~1159 stars 
were made with the boron filter.  These X-ray sources 
will be absent in the WGACAT, and it is necessary to search 
the $ROSAT$ archive for pointed and serendipitous observations
of PG~1159 stars.
PG~1159 stars are divided into two groups: with and without
planetary nebulae (PNs).  
$ROSAT$ observations of cataloged PNs have been analyzed and 
reported by \citet{GCG00}; PSPC observations of five PNs with 
PG~1159 central stars are available.
Of these five, WD\,0726+133 (in Abell 21; PG~1159) and WD\,2333+301 
(in Jn\,1; DOZ.3) are not detected;  WD\,2117+342J (in MWP\,1; DO) 
is detected at a 40$'$ off-axis position in the PSPC field-of-view 
(this paper), but the poor PSF does not allow us to assess accurately
whether faint hard X-ray emission exists; WD\,0044$-$121 (in NGC\,246;
PG~1159) and WD\,1821+643 (in K\,1-16; DOZ.4) are centered in pointed
PSPC observations, but only soft ($<$0.5 keV) X-ray emission from 
the white dwarf is detected.
Four PG~1159 stars without PNs have pointed PSPC observations:
WD\,0122$-$753J (= RX\,J0122$-$7521; DO), WD\,1144+004 (= PG~1144+005; 
DQZO1), WD\,1159$-$034 (= PG~1159$-$034; DQZO.4), and WD\,1501+664 
(= PG~1501+661; DZ1); all four have been included in our survey  
listed in Table 1.
Soft X-ray emission is detected in these four PG~1159 stars without
PNs, but 3-$\sigma$ detection of hard X-rays is obtained only
for PG~1159 itself.

We have further combed Table 1 for hot white dwarfs similar to 
WD\,0005+511 or WD\,2226$-$210, and find a hot DO white dwarf, 
WD\,1522+662, and a hot DAO white dwarf, WD\,1957+225 
at the center of the Dumbbell Nebula.
These four hot DO and DAO white dwarfs and the above nine PG~1159 
white dwarfs will be discuss below in more detail.  
The spectral type, stellar effective temperature, visual magnitude, 
and $ROSAT$ observations of these 13 hot white dwarfs are summarized
in Table 4.
To illustrate the spectral properties of the 11 hot white dwarfs 
whose X-ray emission has been detected by $ROSAT$ PSPC observations,
in Figure 4 we present their soft X-ray images in the 0.1--0.4 keV 
band and the hard X-ray images in the 0.6--2.4 keV band.
It is evident that hard X-ray emission is clearly detected only
from WD\,0005+511, WD\,1159$-$034, and WD\,2226$-$210, as we have
concluded earlier, and possibly detected at a 2$\sigma$ level 
from WD\,0122$-$753J.
The spectral types of these four white dwarfs are DO, DQZO.4, DAO,
and DO, respectively.  

Below we divide the 11 hot white dwarfs detected in X-rays according 
to their spectral types, discuss the implications of their hard X-ray 
properties, and speculate on the possibility of a photospheric origin 
of the hard X-ray emission.

\begin{itemize}
\item {\it DAO (WD\,1957+225 and WD\,2226$-$210):} \\
Hard X-ray emission is detected from WD\,2226$-$210 in the Helix 
Nebula, but not from WD\,1957+225 in the Dumbbell Nebula.
As these two DAO white dwarfs have similar effective temperatures,
and as WD\,1957+225 is only 0.8 mag fainter but has a four times longer
exposure time, the lack of hard X-ray emission from WD\,1957+225 
signifies a real difference from WD\,2226$-$210. 
Furthermore, WD\,2226$-$210 shows temporal variations of its hard X-ray 
emission and H$\alpha$ line profile, suggesting the existence 
of a late dMe companion \citep{GUE01,GRU01}.  
Therefore, there is no evidence indicating that deep photospheric
emission is the origin of hard X-rays associated with DAO white dwarfs.
\item {\it DO (WD\,0005+511, WD\,0122$-$753J, WD\,1522+662, and
  WD\,2117+342J):} \\
Among these four DO white dwarfs, WD\,0005+511 has an 
unambiguous detection of hard X-rays, and WD\,0122$-$753J has
a possible detection of hard X-rays.  
WD\,1522+662 does not show hard X-ray emission, but the $ROSAT$ 
observation has only 4.7 ks exposure time and WD\,1522+662 is faint
with $B$ = 16.4 mag.
WD\,2117+342J is observed at a 40$'$ from the center 
of the PSPC field-of-view and the poor PSF prohibits a conclusive 
assessment of the existence of faint hard X-ray emission.
It may be possible that faint hard X-ray emission can emerge from 
the hot, deep layer of the photospheres of DO white dwarfs, but more
detections are needed to confirm it.
\item {\it DQZO (WD\,1144+044, WD\,1159$-$034, and WD\,1821+643):} \\
WD\,1159$-$034 has hard X-ray emission.  WD\,1144+044 has a 
non-detection, but it is slightly fainter than WD\,1159$-$034
and its exposure time is only half as long.
WD\,1821+643 in the PN K\,1-16 is projected near a bright hard 
X-ray source, a cluster of galaxies surrounding the QSO E1821+643 
\citep{Sax97}, so it is difficult to determine its hard X-ray 
properties accurately from Figure 4.
We have examined an archival $Chandra$ HETG observation of 
QSO E1821+643 (PI: C.\ R.\ Canizares; 101 ks); in this observation
WD\,1821+643 is detected and clearly resolved from QSO E1821+643, 
but no hard X-ray emission from WD\,1821+643 is seen.
WD\,1159$-$034 and WD\,1821+643 are well-known pulsators
\citep{CB96}, while WD\,1144+044 is not.
It is thus unlikely that the hard X-ray emission is related to
the stellar pulsation.
\item {\it PG~1159 and DZ1 (WD\,0044$-$121 and WD\,1501+664):}  \\
WD\,0044$-$121 is a PG~1159 star in the PN NGC\,246;
WD\,1501+664 is classified as DZ1 and is the hottest white dwarf
\citep{WW99}.  
Neither of these two hot white dwarfs show hard X-ray emission; 
the high opacity of the H- and He-free atmosphere of 
WD\,1501+664 may be the culprit of its non-detection 
\citep{Netal86,We91}.
\end{itemize}

The above sample of hot white dwarfs is very limited, but allows
us to eliminate improbable origins of hard X-ray emission.
For example, stellar pulsation is not likely to produce hard
X-ray emission, as WD\,1159$-$034 is the only pulsator with
hard X-ray emission.  Stellar winds are not likely to be the
origin of hard X-ray emission either, as WD\,1159$-$034 has 
hard X-rays but no measurable past or ongoing mass loss 
\citep{FLS90}, while appreciable mass loss but no hard X-ray
emission has been detected from WD\,0044$-$121 and
WD\,1821+643 \citep{KW98}.
There leaves the tantalizing suggestion that hard X-rays may 
be emitted by hot DO and DQZO white dwarfs.  
While it is necessary to model their atmospheres to understand
theoretically whether hard X-rays from beneath the atmosphere
may leak through, it is also necessary to obtain better
high-resolution X-ray images to confirm that the hard X-ray 
emission is indeed associated with the white dwarfs (as opposed 
to background objects projected in their vicinity) and 
high-quality X-ray spectra for detailed spectral analysis.

\section{Summary}

We have correlated the recent white dwarf catalog by 
\citet{MCC99} with the $ROSAT$ PSPC point source catalog 
(WGACAT), and found 76 white dwarfs coincident with X-ray
sources at a high level of confidence.  We have further 
found that 17 of these sources show hard X-ray emission 
($>$ 0.5 keV). Two of these white dwarfs with hard X-ray 
emission accrete material from their companions, while the 
other ten have late-type companions that are known to have 
active coronae and emit hard X-rays.  One white dwarf
has an AGN projected within the PSPC PSF contributing to
the hard X-ray emission.  The remaining four white dwarfs 
(WD\,0339$-$451, WD\,1159$-$034, WD\,1234+481, and WD\,1333+510)
and two additional white dwarfs (WD\,0005+511 and
WD\,2226$-$210) with hard X-ray emission appear single.  
The lack of near-IR excess for WD\,1234+481 and WD\,2226$-$210
at known distances constrains the possible spectral types of 
the hidden companions to later than M7\,V and M5-6\,V,
respectively.
We suggest that hard X-ray emission may be more effective than
near-IR photometry in diagnosing faint, late-type companions
of white dwarfs, if the hard X-ray emission associated with 
the six apparently single white dwarfs indeed originates from
hidden companions.

It is intriguing that three of the six apparently single 
white dwarfs with hard X-ray emission have high stellar 
effective temperatures.  We have searched the $ROSAT$ archive
for observations of PG~1159 stars and examined the X-ray properties
of a sample of 13 hot white dwarfs with different spectral types.
Comparisons among these hot white dwarfs lead to the following 
conclusions: (1) DAO WD\,2226$-$210 most likely possesses a late
dMe companion which emits hard X-rays,
(2) stellar pulsation cannot be connected to the hard X-ray
emission, (3) fast stellar winds are not likely to be the origin
of hard X-ray emission, and (4) the high-energy Wien tail of 
emission deep in the atmospheres of hot DO 
and DQZO white dwarfs remains a tantalizing explanation for
the hard X-ray emission observed.

Our statistics are limited by the incompleteness of the WGACAT, 
which has been derived solely from pointed $ROSAT$ PSPC observations 
made without filters, while many PSPC observations of white dwarfs 
were made with a boron filter. 
A complete survey for white dwarfs with hard X-ray emission 
using the entire $ROSAT$ archive is needed to enlarge the 
sample.
Follow-up high-resolution deep X-ray observations with $Chandra$ 
or {\it XMM-Newton} are needed to confirm the positional
coincidence of the white dwarfs and the X-ray sources, and 
to study the spectral properties in order to investigate 
the origin and nature of hard X-ray emission associated 
with white dwarfs.

\acknowledgments
We thank the anonymous referee for making critical comments
which helped improve our paper.  
We also thank J.\ Liebert, R.\ Napiwotzki, R.\ Petre, and K.\ 
Werner for reading the manuscript and making useful suggestions.
This research has made use of the SIMBAD database, operated at 
CDS, Strasbourg, France, and the Digital Sky Survey produced at 
the Space Telescope Science Institute under U.S.\ Government grant 
NAG W-2166. We have also used data product from the 2MASS, 
which is a joint project of the University Massachusetts and 
the Infrared Processing and Analysis Center/California 
Institute of Technology, funded by NASA and NSF.

%\end{document}

\newpage
%\begin{center}
\begin{figure}
\centerline{\bf Figure Captions}
\caption{WD\,1910+047.  Left: Broad-band Digitized Sky Survey 
image overlaid by X-ray contours at 10, 20, 50, 70, and 90\% of 
the peak value of the X-ray source.  The white dwarf is marked
by two short lines near the center of the field of view.
The X-ray source is clearly associated 
with a star 80$''$ SE of the white dwarf. Right: $ROSAT$ PSPC 
spectrum of the X-ray source.  The X-ray contours and spectrum 
are extracted from the 20.6 ks observation rp500058a02.
}
\end{figure}
%\end{center}

\newpage
\begin{figure}
\caption{Seventeen white dwarfs associated with hard X-ray emission.
Left panels: Broad-band Digitized Sky Survey image overlaid by 
X-ray contours at 10, 20, 50, 70, and 90\% of the peak value of the 
X-ray source.  
In some cases, the background is so high that the lower X-ray 
contours are absent.
The white dwarf is at the center of each field of view.
When multiple stars are projected 
in the vicinity of a white dwarf, we mark the white dwarf with 
two short lines.  
For white dwarfs with large proper motions, lines are drawn
to show the proper motion from the optical epoch to the X-ray
epoch.
For WD\,0736+053, the wavy line shows the proper and orbital
motions of the white dwarf; for WD\,1633+572, the arrow shows
the motion of its common proper motion companion CM Dra.
Right panels: $ROSAT$ PSPC spectrum of the X-ray source.  In cases 
where the spectrum is overwhelmed by soft X-ray emission, to show the 
hard X-ray emission, the spectrum is also plotted with open triangles
after being multiplied by a constant factor indicated in the panel.
The X-ray contours and spectra are extracted from $ROSAT$ PSPC
observations listed in Table 1.
}
\end{figure}

%\begin{figure}
%\psfig{file=fig3.ps,width=16cm}
%\caption{Soft- and hard-band X-ray images of WD\,1159$-$034 and 
%WD\,0548+000.  Left panels: soft-band in the 0.1--0.5 keV energy
%range.  Right panels: hard-band in the 0.9--2.4 keV energy range.
%The soft-band images detect similar number of counts for 
%each source, while the hard-band images show weak emission only
%from WD\,1159$-$034.
%}

\begin{figure}
\caption{KPD~0005+5106 (= WD\,0005+511).  Left: Broad-band Digitized 
Sky Survey image overlaid by X-ray contours at 10, 20, 50, 70, and 
90\% of the peak value of the X-ray source.  The white dwarf is marked
by two short lines near the center of the field of view.  
Two X-ray sources associated with two different stars are detected.
Right: $ROSAT$ PSPC spectrum of the X-ray source associated with the
white dwarf KPD~0005+5106.  This spectrum was carefully extracted to 
exclude the X-ray emission from the neighboring source.
The X-ray contours and spectrum are extracted from the 5 ks 
observation rf200428n00 made with a boron filter.  Hard X-ray 
emission at 1 keV is clearly detected.
}
\end{figure}

\begin{figure}
\caption{
Soft and hard X-ray images of eleven apparently single 
hot white dwarfs.  The soft X-ray images (left panels)
are extracted from PSPC observations in the 0.1--0.4 keV 
energy band, and the hard X-ray images (right panels)
in the 0.6--2.4 keV energy band.
}
\end{figure}

\newpage
 
{\setlength{\textheight}{9.0in}     % height of main text
\setlength{\textwidth}{6.5in}        % width of text
\renewcommand{\arraystretch}{.6}
\begin{deluxetable}{lllllrrrrr}
\tablewidth{0pt}
\rotate
%\tabletypesize{\scriptsize} 
%\tablenum{1}
\tablecaption{White Dwarfs Coincident with WGACAT X-ray Sources}
\tablehead{
~~~~(1) & ~~~~(2) & ~~(3) & ~~~~~~(4) & ~~~(5) & (6)~ & (7)~ & \multicolumn{1}{c}{(8)} & \multicolumn{1}{c}{(9)} & \multicolumn{1}{c}{(10)} \\
       &        &      &             & ROSAT     & Exp.  &      & \multicolumn{3}{c}{X-ray Counts Reported in WGACAT}      \\
WD     & Common & WD   & WGACAT      & ~PSPC     & Time  & Pos. & \multicolumn{1}{c}{0.1-0.4 kev}& \multicolumn{1}{c}{0.4-0.9 kev} & \multicolumn{1}{c}{0.9-2.4 kev} \\
Number & Name   & Type & Number      & Obs.\ \#  & [ks]~  & Coin. & \multicolumn{1}{c}{[cts]}      & \multicolumn{1}{c}{[cts]}       & \multicolumn{1}{c}{[cts]}     
}
\startdata
0027$-$636 &  ...                   &  DA1       & J0029.9$-$6324 & 400160  &  2.7 & good &    4081 $\pm$ 64\phn\phn    &     27 $\pm$ 5\phn\phn\phn &      0 $\pm$ 0\phn\phn\phn \\ 
0037$+$312 &  GD\,8                 &  DA1       & J0039.8$+$3132 & 201045  & 28.4 & good &    2248 $\pm$ 47\phn\phn    &    106 $\pm$ 10\phn\phn    &     65 $\pm$ 8\phn\phn\phn \\ 
0048$-$294 &  FOCAP\,SGP2:31        &  DA        & J0051.2$-$2910 & 700275  & 24.5 & U1,2 &      46 $\pm$ 7\phn\phn\phn &     11 $\pm$ 3\phn\phn\phn &     24 $\pm$ 5\phn\phn\phn \\ 
0050$-$332 &  GD\,659               &  DA1       & J0053.2$-$3300 & 200410  &  3.4 & good &    3787 $\pm$ 62\phn\phn    &     27 $\pm$ 5\phn\phn\phn &      3 $\pm$ 2\phn\phn\phn \\ 
0116$-$231 &  GD\,695               &  DA3       & J0118.6$-$2254 & 100376  & 17.0 & good &      50 $\pm$ 7\phn\phn\phn &      5 $\pm$ 2\phn\phn\phn &      1 $\pm$ 1\phn\phn\phn \\ 
0122$-$753J&                        &  DO        & J0122.8$-$7521 & 300369  &  5.6 & good &    3130 $\pm$ 56\phn\phn    &     98 $\pm$ 10\phn\phn    &      8 $\pm$ 3\phn\phn\phn \\  
0131$-$163 &  GD\,984               &  DA1$+$dM  & J0134.4$-$1607 & 200485  &  0.9 & good &     633 $\pm$ 25\phn\phn    &      4 $\pm$ 2\phn\phn\phn &      0 $\pm$ 0\phn\phn\phn \\ 
0136$+$251 &  PG\,0136$+$251        &  DAp1      & J0138.8$+$2523 & 200539  &  2.1 & good &     550 $\pm$ 24\phn\phn    &      5 $\pm$ 2\phn\phn\phn &      0 $\pm$ 0\phn\phn\phn \\ 
0216$-$032 &  VZ Cet                &  DA+M2-7III& J0219.3$-$0258 & 201501  &  9.1 & good &       5 $\pm$ 2\phn\phn\phn &     18 $\pm$ 4\phn\phn\phn &     40 $\pm$ 6\phn\phn\phn \\ 
0304$+$154 &  ...                   &  DC        & J0307.0$+$1540 & 800104  &  7.7 & U1,2 &      48 $\pm$ 7\phn\phn\phn &     11 $\pm$ 3\phn\phn\phn &      5 $\pm$ 2\phn\phn\phn \\ 
0320$-$539 &  LB\,1663              &  DA1.5     & J0322.2$-$5345 & 800307  & 21.5 & good &    2534 $\pm$ 50\phn\phn    &     14 $\pm$ 3\phn\phn\phn &      1 $\pm$ 1\phn\phn\phn \\ 
0333$-$350 &  ...                   &  DA        & J0335.5$-$3449 & 600127  & 18.0 & good &    1480 $\pm$ 38\phn\phn    &     20 $\pm$ 4\phn\phn\phn &      5 $\pm$ 2\phn\phn\phn \\ 
0339$-$451 &  ...                   &  DA        & J0341.4$-$4500 & 900495  & 48.6 & good &      64 $\pm$ 8\phn\phn\phn &     22 $\pm$ 5\phn\phn\phn &     18 $\pm$ 4\phn\phn\phn \\ 
0347$+$171 &  V471\,Tau             &  DA2$+$K2V & J0350.4$+$1714 & 200107  & 31.8 & good &    6538 $\pm$ 81\phn\phn    &    688 $\pm$ 26\phn\phn    &    809 $\pm$ 28\phn\phn   \\ 
0416$-$550 &  ...                   &  DA        & J0417.1$-$5457 & 600456  & 17.8 & good &     254 $\pm$ 16\phn\phn    &     16 $\pm$ 4\phn\phn\phn &     11 $\pm$ 3\phn\phn\phn \\ 
0425$+$168 &  EGGR\,37              &  DA2       & J0428.6$+$1658 & 200083  &  2.9 & good &      33 $\pm$ 6\phn\phn\phn &      3 $\pm$ 2\phn\phn\phn &      0 $\pm$ 0\phn\phn\phn \\ 
0426$+$588 &  GJ\,169.1B            &  DC$+$M4   & J0431.1$+$5859 & 201114  &  3.5 & good &      22 $\pm$ 5\phn\phn\phn &      4 $\pm$ 2\phn\phn\phn &      4 $\pm$ 2\phn\phn\phn \\ 
0429$+$176 &  HZ\,9                 &DA2$+$dM4.5e& J0432.4$+$1744 & 200443  & 20.3 & good &     384 $\pm$ 20\phn\phn    &    180 $\pm$ 13\phn\phn    &    161 $\pm$ 13\phn\phn   \\ 
0443$-$037J&  RE\,J0443$-$034       &  DA        & J0443.0$-$0346 & 200997  & 10.1 & good &   14710 $\pm$ 121\phn       &    336 $\pm$ 18\phn\phn    &     42 $\pm$ 6\phn\phn\phn \\
0446$-$789 &  ...                   &  DA3       & J0443.6$-$7851 & 201073  &  8.3 & good &      74 $\pm$ 9\phn\phn\phn &      1 $\pm$ 1\phn\phn\phn &      1 $\pm$ 1\phn\phn\phn \\ 
0518$-$105 &  RE\,J0521$-$102       &  DA2       & J0521.3$-$1029 & 200830  &  5.3 & good &     370 $\pm$ 19\phn\phn    &      1 $\pm$ 1\phn\phn\phn &      0 $\pm$ 0\phn\phn\phn \\ 
0531$-$022 &  RE\,J0534$-$021       &  DA2       & J0534.3$-$0213 & 200932  &  8.0 & good &     284 $\pm$ 17\phn\phn    &     17 $\pm$ 4\phn\phn\phn &      6 $\pm$ 2\phn\phn\phn \\ 
0548$+$000 &  GD\,257               &  DA1       & J0550.6$+$0005 & 200585  &  4.4 & good &    1404 $\pm$ 38\phn\phn    &     18 $\pm$ 4\phn\phn\phn &      0 $\pm$ 0\phn\phn\phn \\ 
0558$-$756 &  ...                   &  DO        & J0556.9$-$7540 & 201245  & 18.5 & U2,3 &      10 $\pm$ 3\phn\phn\phn &      5 $\pm$ 2\phn\phn\phn &     13 $\pm$ 4\phn\phn\phn \\ 
0651$-$020 &  GD\,80                &  DA1       & J0654.2$-$0209 & 201363  &  0.4 & good &      70 $\pm$ 8\phn\phn\phn &      1 $\pm$ 1\phn\phn\phn &      0 $\pm$ 0\phn\phn\phn \\ 
0715$-$704J&  RE\,J0715-705         &  DA1       & J0715.2$-$7025 & 400321  &  2.8 & good &    5439 $\pm$ 74\phn\phn    &     27 $\pm$ 5\phn\phn\phn &      0 $\pm$ 0\phn\phn\phn \\ 
0718$-$316 &  IN\,CMa               &  DAO$+$dM  & J0720.7$-$3146 & 300338  &  3.2 & good &     411 $\pm$ 20\phn\phn    &      4 $\pm$ 2\phn\phn\phn &      2 $\pm$ 1\phn\phn\phn \\ 
0734$-$143 &  ...                   &  DA        & J0736.6$-$1428 & 500295  &  9.1 & U3   &       5 $\pm$ 2\phn\phn\phn &     19 $\pm$ 4\phn\phn\phn &     23 $\pm$ 5\phn\phn\phn \\ 
0736$+$053 &  Procyon B             &  DA+F5IV-V & J0739.3$+$0513 & 200437  &  3.8 & good &   11700 $\pm$ 110\phn       &    874 $\pm$ 30\phn\phn    &    118 $\pm$ 11\phn\phn   \\ 
0800$-$477J&                        &  DA        & J0800.4$-$4745 & 400158  &  2.4 & good &    1333 $\pm$ 37\phn\phn    &     42 $\pm$ 6\phn\phn\phn &      0 $\pm$ 0\phn\phn\phn \\
0805$+$654 &  PG\,0805$+$654        &  DA1       & J0809.6$+$6518 & 700258  &  8.4 & U2   &     198 $\pm$ 14\phn\phn    &     67 $\pm$ 8\phn\phn\phn &     18 $\pm$ 4\phn\phn\phn \\ 
0824$+$288 &  PG0824$+$289          &DA$+$dC$+$M3& J0827.0$+$2844 & 201083  &  2.7 & good &     586 $\pm$ 24\phn\phn    &      5 $\pm$ 2\phn\phn\phn &      1 $\pm$ 1\phn\phn\phn \\ 
0839$-$528 &  IC\,2391~KR\,1        &  DA3       & J0841.1$-$5300 & 200501  & 22.9 & U2,3 &      16 $\pm$ 4\phn\phn\phn &     14 $\pm$ 4\phn\phn\phn &     30 $\pm$ 6\phn\phn\phn \\ 
0840$+$200 &  LB\,1876              &  DA5       & J0842.8$+$1951 & 200250  &  1.9 & U2   &      11 $\pm$ 3\phn\phn\phn &      4 $\pm$ 2\phn\phn\phn &     10 $\pm$ 3\phn\phn\phn \\
0841$+$033J&   RE\,J0841$+$032      &  DA1       & J0841.0$+$0320 & 201362  &  0.3 & good &     182 $\pm$ 13\phn\phn    &      1 $\pm$ 1\phn\phn\phn &      0 $\pm$ 0\phn\phn\phn \\ 
0842$+$490 &  HD\,74389B            &  DA$+$A2V  & J0845.8$+$4852 & 200816  &  0.3 & good &      91 $\pm$ 10\phn\phn    &      3 $\pm$ 2\phn\phn\phn &      0 $\pm$ 0\phn\phn\phn \\ 
0903$+$166 &  ...                   &  ...       & J0905.9$+$1624 & 700385  &  6.2 & U2,3 &      21 $\pm$ 5\phn\phn\phn &     21 $\pm$ 5\phn\phn\phn &     12 $\pm$ 3\phn\phn\phn \\ 
0904$+$511 &  PG\,0904$+$511        &  DA1.5     & J0907.7$+$5058 & 800474  &  9.6 & good &     524 $\pm$ 23\phn\phn    &     25 $\pm$ 5\phn\phn\phn &     26 $\pm$ 5\phn\phn\phn \\ 
0916$-$197J&  RE\,J0916$-$194       &  DA        & J0916.9$-$1946 & 400162  &  2.9 & good &    1494 $\pm$ 39\phn\phn    &     19 $\pm$ 4\phn\phn\phn &      0 $\pm$ 0\phn\phn\phn \\ 
0937$+$505 &  PG\,0937$+$505        &  DA1       & J0940.3$+$5021 & 200957  &  4.1 & good &     239 $\pm$ 16\phn\phn    &      0 $\pm$ 0\phn\phn\phn &      0 $\pm$ 0\phn\phn\phn \\ 
0954$+$697 &  PG\,0954$+$697        &  DA2.5     & J0958.4$+$6928 & 600101  & 21.4 & U4   &     174 $\pm$ 13\phn\phn    &     76 $\pm$ 9\phn\phn\phn &     38 $\pm$ 6\phn\phn\phn \\ 
1010$+$064 &  PG\,1010$+$065        &  DA1       & J1013.4$+$0612 & 200540  &  6.8 & good &     204 $\pm$ 14\phn\phn    &      5 $\pm$ 2\phn\phn\phn &      2 $\pm$ 1\phn\phn\phn \\ 
1032$+$534J&  RE\,J1032$+$535       &  DA1       & J1032.1$+$5330 & 900149  & 17.7 & good &   40710 $\pm$ 202\phn       &    335 $\pm$ 18\phn\phn    &     81 $\pm$ 9\phn\phn\phn \\     
1040$+$451 &  PG\,1040$+$451        &  DA1       & J1043.5$+$4454 & 201020  & 14.7 & good &    1787 $\pm$ 42\phn\phn    &     66 $\pm$ 8\phn\phn\phn &     28 $\pm$ 5\phn\phn\phn \\ 
1059$+$514J&                        &  DA        & J1059.2$+$5124 & 400159  &  3.2 & good &    6582 $\pm$ 81\phn\phn    &    102 $\pm$ 10\phn\phn    &      0 $\pm$ 0\phn\phn\phn \\
1109$+$244 & PG\,1109$+$244         &  DA1.5     & J1112.6$+$2409 & 201365  &  0.2 & good &      46 $\pm$ 7\phn\phn\phn &      0 $\pm$ 0\phn\phn\phn &      0 $\pm$ 0\phn\phn\phn \\ 
1134$+$300 &  GJ\,433.1             &  DA2       & J1137.0$+$2948 & 200091  & 33.9 & good &     327 $\pm$ 18\phn\phn    &     59 $\pm$ 8\phn\phn\phn &     51 $\pm$ 7\phn\phn\phn \\ 
1144$+$004 &  ...                   &  DQZO1     & J1146.5$+$0012 & 201242  &  5.8 & good &     149 $\pm$ 12\phn\phn    &      5 $\pm$ 2\phn\phn\phn &      3 $\pm$ 2\phn\phn\phn \\ 
1159$-$034 &  GW\,Vir               &  DQZO.4    & J1201.7$-$0345 & 701202  & 13.6 & good &     779 $\pm$ 28\phn\phn    &     12 $\pm$ 4\phn\phn\phn &     13 $\pm$ 4\phn\phn\phn \\ 
1213$+$528 &  EG\,UMa               &  DA4$+$dM2e& J1215.6$+$5230 & 200953  &  2.6 & good &     249 $\pm$ 16\phn\phn    &     68 $\pm$ 8\phn\phn\phn &     84 $\pm$ 9\phn\phn\phn \\ 
1229$+$290 &  ...                   &  DC        & J1231.7$+$2848 & 201163  &  1.8 & U1,3 &     113 $\pm$ 11\phn\phn    &     44 $\pm$ 7\phn\phn\phn &     39 $\pm$ 6\phn\phn\phn \\ 
1234$+$481 &  PG\,1234$+$481        &  DA1       & J1236.7$+$4755 & 200578  &  2.5 & good &    2127 $\pm$ 46\phn\phn    &     89 $\pm$ 9\phn\phn\phn &      0 $\pm$ 0\phn\phn\phn \\ 
1254$+$223 &  GD\,153               &  DA1       & J1257.0$+$2201 & 132471  &  8.6 & good &   16673 $\pm$ 130\phn       &     95 $\pm$ 10\phn\phn    &      0 $\pm$ 0\phn\phn\phn \\ 
1255$+$258J&  HD\,112313            & CSPN+G5III & J1255.5$+$2553 & 201514  & 18.8 & good &      90 $\pm$ 9\phn\phn\phn &     29 $\pm$ 5\phn\phn\phn &     39 $\pm$ 6\phn\phn\phn \\
1314$+$293 &  HZ\,43A               &DA1$+$dM3.5e& J1316.3$+$2906 & 100308  & 21.5 & good & 1585270 $\pm$ 1260          &    7312 $\pm$ 86\phn\phn   &    120 $\pm$ 11\phn\phn   \\ 
1317$+$453 &  GJ\,2100              &  DA3.5     & J1319.2$+$4505 & 900325  & 10.3 & U1,2 &      34 $\pm$ 6\phn\phn\phn &     12 $\pm$ 4\phn\phn\phn &     14 $\pm$ 4\phn\phn\phn \\ 
1325$+$581 &  EGGR\,358             &  DA7       & J1327.6$+$5755 & 600458  & 18.1 & U2,4 &     247 $\pm$ 16\phn\phn    &     35 $\pm$ 6\phn\phn\phn &     44 $\pm$ 7\phn\phn\phn \\ 
1333$+$510 &  PG\,1333$+$510        &  DA        & J1335.2$+$5049 & 800047  & 16.4 & good &     447 $\pm$ 21\phn\phn    &     51 $\pm$ 7\phn\phn\phn &     47 $\pm$ 7\phn\phn\phn \\ 
1403$-$077 &  PG\,1403$-$077        &  DA1       & J1406.0$-$0758 & 200528  &  4.8 & good &     345 $\pm$ 19\phn\phn    &      9 $\pm$ 3\phn\phn\phn &      1 $\pm$ 1\phn\phn\phn \\ 
1446$+$634J&  ...                   &  DA1       & J1446.0$+$6329 & 700975  &  4.2 & good &     467 $\pm$ 22\phn\phn    &      5 $\pm$ 2\phn\phn\phn &      2 $\pm$ 1\phn\phn\phn \\ 
1501$+$664 &  RE\,J1502$+$661       &  DZ1       & J1502.1$+$6612 & 170001  & 43.2 & good &  357416 $\pm$ 600\phn       &   1689 $\pm$ 41\phn\phn    &     12 $\pm$ 4\phn\phn\phn \\ 
1522$+$662 &  ...                   &  DO        & J1522.9$+$6604 & 201240  &  4.7 & good &     742 $\pm$ 27\phn\phn    &     13 $\pm$ 4\phn\phn\phn &      1 $\pm$ 1\phn\phn\phn \\ 
1620$-$391 &  EGGR\,274             &  DA2       & J1623.5$-$3913 & 200588  &  1.9 & good &     634 $\pm$ 25\phn\phn    &      2 $\pm$ 1\phn\phn\phn &      2 $\pm$ 1\phn\phn\phn \\ 
1631$+$781 &  1ES\,1631$+$78.1      &  DA1$+$dM4e& J1629.1$+$7804 & 170154  & 37.5 & good &  125636 $\pm$ 350\phn       &   1355 $\pm$ 37\phn\phn    &    350 $\pm$ 19\phn\phn   \\ 
1633$+$572 &  GJ\,630.1B            & DQ8$+$dM4e & J1634.3$+$5709 & 200721  & 47.5 & good &    2397 $\pm$ 49\phn\phn    &   1249 $\pm$ 35\phn\phn    &   1038 $\pm$ 32\phn\phn   \\ 
1634$-$573 &  HD\,149499B           & DOZ1$+$K2Ve& J1638.5$-$5728 & 200773  &  1.4 & good &     240 $\pm$ 16\phn\phn    &    145 $\pm$ 12\phn\phn    &    146 $\pm$ 12\phn\phn   \\ 
1636$+$351 &  PG\,1636$+$351        &  DA1.5     & J1638.4$+$3500 & 201082  &  4.2 & good &     909 $\pm$ 30\phn\phn    &      9 $\pm$ 3\phn\phn\phn &      0 $\pm$ 0\phn\phn\phn \\ 
1641$+$387 &  GD\,357               &  DA3       & J1643.1$+$3840 & 201538  &  5.5 & U1   &      26 $\pm$ 5\phn\phn\phn &     10 $\pm$ 3\phn\phn\phn &     13 $\pm$ 4\phn\phn\phn \\ 
1650$+$406J&  RE\,J1650$+$403       &  DA1       & J1650.3$+$4037 & 201080  &  2.2 & good &      55 $\pm$ 7\phn\phn\phn &      1 $\pm$ 1\phn\phn\phn &      0 $\pm$ 0\phn\phn\phn \\ 
1657$+$343 &  PG\,1657$+$343        &  DA2       & J1658.8$+$3418 & 201079  &  6.4 & good &     591 $\pm$ 24\phn\phn    &      7 $\pm$ 3\phn\phn\phn &      0 $\pm$ 0\phn\phn\phn \\ 
1658$+$440 &  PG\,1658$+$440        &  DAp1      & J1659.8$+$4400 & 201078  &  6.7 & good &     503 $\pm$ 22\phn\phn    &      7 $\pm$ 3\phn\phn\phn &      1 $\pm$ 1\phn\phn\phn \\ 
1659$+$442 &  PG\,1659$+$442        &  DA        & J1700.6$+$4410 & 201078  &  6.7 & U1,2 &      22 $\pm$ 5\phn\phn\phn &      7 $\pm$ 3\phn\phn\phn &      7 $\pm$ 3\phn\phn\phn \\ 
1802$+$213 &  GD\,372               &  DA4       & J1804.4$+$2120 & 200940  & 13.7 & U2,3 &      35 $\pm$ 6\phn\phn\phn &     21 $\pm$ 5\phn\phn\phn &     14 $\pm$ 4\phn\phn\phn \\ 
1821$+$643 &  DS\,Dra               &  DOZ.4     & J1821.8$+$6422 & 700948  &  2.0 & good &     135 $\pm$ 12\phn\phn    &      9 $\pm$ 3\phn\phn\phn &     20 $\pm$ 5\phn\phn\phn \\ 
1844$-$654 &  ...                   &  DA        & J1848.9$-$6525 & 200941  & 16.0 & good &     757 $\pm$ 28\phn\phn    &    126 $\pm$ 11\phn\phn    &     64 $\pm$ 8\phn\phn\phn \\ 
1906$-$600 &  ...                   &  DC        & J1910.8$-$5958 & 300047  &  5.2 & U3   &       7 $\pm$ 3\phn\phn\phn &     20 $\pm$ 5\phn\phn\phn &     41 $\pm$ 6\phn\phn\phn \\ 
1910$+$047 &  ...                   &  DA2       & J1912.5$+$0452 & 400271  &  2.8 & U4   &      20 $\pm$ 45\phn\phn    &     10 $\pm$ 3\phn\phn\phn &     27 $\pm$ 5\phn\phn\phn \\ 
1944$-$421 &  V3885\,Sgr            &  CV        & J1947.6$-$4200 & 300232  & 10.4 & good &     345 $\pm$ 19\phn\phn    &    448 $\pm$ 21\phn\phn    &    699 $\pm$ 26\phn\phn   \\ 
1957$+$225 &  ...                   &  DAO       & J1959.6$+$2243 & 900016  &  5.9 & good &     828 $\pm$ 29\phn\phn    &      5 $\pm$ 2\phn\phn\phn &      0 $\pm$ 0\phn\phn\phn \\ 
2013$+$400J&  RE\,J2013$+$400       &  DAO       & J2013.1$+$4002 & 400157  &  3.0 & good &    1166 $\pm$ 34\phn\phn    &     17 $\pm$ 4\phn\phn\phn &      9 $\pm$ 3\phn\phn\phn \\
2014$-$575 &  RE\,J2018$-$572       &  DA2       & J2018.8$-$5721 & 200580  &  4.1 & good &     148 $\pm$ 12\phn\phn    &      3 $\pm$ 2\phn\phn\phn &      0 $\pm$ 0\phn\phn\phn \\ 
2020$-$425 &  UVE\,J2024$-$42.4     &  DA        & J2023.9$-$4224 & 200488  &  0.9 & good &      71 $\pm$ 8\phn\phn\phn &      2 $\pm$ 1\phn\phn\phn &      0 $\pm$ 0\phn\phn\phn \\ 
2028$+$390 &  GD\,391               &  DA2       & J2029.9$+$3913 & 200412  &  2.2 & good &      46 $\pm$ 7\phn\phn\phn &      0 $\pm$ 0\phn\phn\phn &      0 $\pm$ 0\phn\phn\phn \\ 
2032$+$248 &  HD\,340611            &  DA2.5     & J2034.3$+$2504 & 200087  &  9.9 & good &     100 $\pm$ 10\phn\phn    &      2 $\pm$ 1\phn\phn\phn &      5 $\pm$ 2\phn\phn\phn \\
2034$-$275J&                        &  DA        & J2034.9$-$2734 & 201236  &  5.0 & good &     574 $\pm$ 24\phn\phn    &      4 $\pm$ 2\phn\phn\phn &      2 $\pm$ 1\phn\phn\phn \\ 
2052$+$466J&                        &  DO        & J2052.6$+$4639 & 200114  &  3.5 & U3   &     801 $\pm$ 28\phn\phn    &    118 $\pm$ 11\phn\phn    &     27 $\pm$ 5\phn\phn\phn \\ 
2056$+$033 &  PG\,2056$+$033        &  DA1       & J2058.7$+$0332 & 200955  &  5.6 & good &      37 $\pm$ 6\phn\phn\phn &      2 $\pm$ 1\phn\phn\phn &      1 $\pm$ 1\phn\phn\phn \\ 
2111$+$498 &  GD\,394               &  DA1.5     & J2112.7$+$5006 & 200427  &  0.4 & good &     207 $\pm$ 14\phn\phn    &      1 $\pm$ 1\phn\phn\phn &      0 $\pm$ 0\phn\phn\phn \\ 
2117$+$342J&  V2027\,Cyg            &  DO        & J2117.1$+$3412 & 201512  & 26.6 & good &    6588 $\pm$ 81\phn\phn    &    194 $\pm$ 14\phn\phn    &     66 $\pm$ 8\phn\phn\phn \\ 
2153$-$419 &  RE\,J2156$-$414       &  DA        & J2156.5$-$4142 & 200487  &  0.7 & good &     397 $\pm$ 20\phn\phn    &      2 $\pm$ 1\phn\phn\phn &      0 $\pm$ 0\phn\phn\phn \\ 
2154$-$512 &  GJ\,841B              &DQ7$+$dM3-5e& J2157.7$-$5059 & 600146  &  5.1 & good &    1851 $\pm$ 43\phn\phn    &    606 $\pm$ 25\phn\phn    &    706 $\pm$ 27\phn\phn   \\ 
2309$+$105 &  GD\,246               &  DA1       & J2312.3$+$1047 & 100578  & 10.3 & good &   36832 $\pm$ 190\phn       &    141 $\pm$ 13\phn\phn    &      3 $\pm$ 2\phn\phn\phn \\ 
2321$-$549 &  RE\,J2324$-$544       &  DA        & J2324.5$-$5441 & 400166  &  4.1 & good &    1964 $\pm$ 44\phn\phn    &     20 $\pm$ 5\phn\phn\phn &      1 $\pm$ 1\phn\phn\phn \\ 
2357$+$296 &  PG\,2357$+$296        &  DA1       & J0000.1$+$2956 & 200535  &  3.8 & good &      36 $\pm$ 6\phn\phn\phn &      2 $\pm$ 1\phn\phn\phn &      1 $\pm$ 1\phn\phn\phn \\ 
\enddata

\end{deluxetable}
}

\newpage

\renewcommand{\arraystretch}{.6}
\begin{deluxetable}{llllll}
\tablewidth{0pt}
%\tabletypesize{\scriptsize} 
\tablecaption{White Dwarfs with Hard X-ray Emission}
\tablehead{
WD   & Spectral &  Parallax\,\tablenotemark{a} &  log $a$\,\tablenotemark{b} &
$kT$\,\tablenotemark{c} & log  $L_{\rm X}$\,\tablenotemark{d}   \\
Number & Type  &   (mas)     &  (AU)    & (keV) &  (ergs~s$^{-1}$)
}
\startdata
0216$-$032  & DA+M2-7III & \phn\phn7.79$\pm$1.07 & \phn2.04: & 0.5  & 29.9  \\
0339$-$451  & DA         & \phn\phn...           & \phn\phn...   &  ...   & ...\\
0347$+$171  & DA2+K2V    & \phn21.37$\pm$1.62    & --1.81 &   0.8  & 29.8 \\
0429$+$176  & DA2+dM4.5e & \phn21.58\,\tablenotemark{e} & --1.91  &   0.6  & 28.6 \\
0736$+$053  & DA+F5IV-V  & 285.93$\pm$0.88       & \phn1.20  & 0.2    & 27.3 \\
1134$+$300  & DA2+(AGN)  & \phn65.28$\pm$3.61    & \phn\phn...   & ... & ... \\
1159$-$034  & DQZO.4     & \phn\phn...           & \phn\phn...   & ...   & ...\\
1213$+$528  & DA4+dM2e   & \phn33.4\phn$\pm$4.1\,\tablenotemark{f}  & --1.89  &  0.8  & 28.8\\
1234$+$481  & DA1        & \phn\phn6.3\,\tablenotemark{g}  & \phn\phn...   &  ...  & ...\\
1255$+$258J & CSPN+G5III & \phn\phn4.70$\pm$0.75\,\tablenotemark{h}  & $>$0.  & 0.7   & 29.3 \\
1314$+$293  & DA1+dM3.5e & \phn31.26$\pm$8.33    & $\ge$1.85 & 0.6   & 28.0 \\
1333$+$510  & DA         & \phn\phn...           & \phn\phn ...  & ...   & ... \\
1631$+$781  & DA1+dM4e   & \phn19.\,\tablenotemark{g} & $>$0. & 0.8   & 28.7 \\
1633$+$572  & DQ8+2(dM3-4e) & \phn68.4\phn$\pm$3.3\,\tablenotemark{i} &$\ge$2.58&  0.4  & 28.0 \\
1634$-$573  & DOZ1+K2Ve  & \phn26.94$\pm$1.88\,\tablenotemark{j} & $\ge$1.69 &  0.7  & 29.3 \\
1944$-$421  & CV         & \phn\phn9.11$\pm$1.95 &  $-$2.09 &  1.2  & 30.3 \\
2154$-$512  & DQ7+2(dM3-5e) & \phn61.63$\pm$2.67\,\tablenotemark{k} & $\ge$2.66 &  0.8  & 29.0  \\
\enddata
\tablenotetext{a}{$Hipparcos$ trigonometric parallaxes \citep{Pe97} unless otherwise noted.}
\tablenotetext{b}{These values are derived from visual orbits, apparent separations, or
  orbital periods cited in \S3.1, supplemented by parallaxes tabulated here and, for short-period 
  systems, mass estimates from \citet{RK98}.  For V471 Tau = WD\,0347+171, masses were adopted from
  \citet{Oetal01}.}
\tablenotetext{c}{For $T$ = 10$^6$ K, $kT$ = 0.086 keV.}
\tablenotetext{d}{In the 0.5 -- 2.4 keV energy band.}
\tablenotetext{e}{Assumes membership in the Hyades \citep{vAl69} 
  with distance from \citet{Pe98}.}
\tablenotetext{f}{Trigonometric parallax from \citet{Detal82}.}
\tablenotetext{g}{Parallax deduced from EUV/IR photometric models of \citet{GAN00}.}
\tablenotetext{h}{Trigonometric parallax from \citet{Hetal97}.}
\tablenotetext{i}{Trigonometric parallax from \citet{Ha80}; weighted mean for both visual components.}
\tablenotetext{j}{$Hipparcos$ parallax of the binary companion  HD\,149499A \citep{We79}.}
\tablenotetext{k}{$Hipparcos$ parallax of the common proper motion binary
companion GL\,841A \citep{JB93}.}
\end{deluxetable}

\renewcommand{\arraystretch}{.6}
\begin{deluxetable}{lllllll}
\tablewidth{0pt}
\tablecaption{Multi-band Photometry of White Dwarfs with Hard X-rays}
\tablehead{ 
WD     &  Spectral  &  $V$  &  $J$  &  $H$  &  $K$  & 
   2MASS\,\tablenotemark{a}   \\
Number &    Type    & (mag) & (mag) & (mag) & (mag) & Source 
}
\startdata 
0005+511 &  DO      & 13.32 & 13.93 & 14.13 & 14.18 & J0008181+512316\\
0216$-$032 & DA+M2-7III & 11.32\,\tablenotemark{b} & --1.06 &
  --1.89  &  --2.44   &  J2192081$-$025841\,\tablenotemark{c} \\
0429+176&DA2+dM4.5e & 13.93 & 10.76 & 10.12 & \phn9.93 & J0432237+174502\\
1134+300 &  DA2+(AGN) & 12.50 & 12.95 & 13.04 & 13.13 & \tablenotemark{d} \\
1159$-$034 & DQZO.4 & 14.87\,\tablenotemark{e}
 & 15.58 & 15.87 & 15.78 & J1201459$-$034540\\
1234+481 & DA1      & 14.42 & 14.99 & 15.07 & 15.12 & J1236451+475522\\
1631+781 & DA1+dM4e & 13\,\tablenotemark{f}
    & 11.00 & 10.28 & 10.15 & J1629102+780439\\
1633+572 &DQ8 +     & 14.99 & 14.09 & 14.08 & 14.07 & J1634216+571008 \\
         & 2(dM3-4e)& 12.87\,\tablenotemark{g} & \phn8.50 & \phn8.04 & 
  \phn7.77 &
         J1634204+570943\\
2226$-$210 & DAO    & 13.54 & 14.35 & 14.50 & 14.62 & J2229385$-$205013\\
\enddata
\tablenotetext{a}{The Two Micron All Sky Survey (2MASS) is a joint 
project of the University of Massachusetts and the Infrared Processing 
and Analysis Center/California Institute of Technology.}
\tablenotetext{b}{Hubble Space Telescope $F550M$ magnitude, equivalent
  to  Str\"omgren $y$ magnitude, of the white dwarf alone
  \citep{Ketal97}. }
\tablenotetext{c}{Saturated in 2MASS survey.  $J$, $H$, and $K$ from
\citet{Cetal79}.  Variable in all bandpasses.}
\tablenotetext{d}{$V$ -- \citet{MG90}; $J$, $H$, and $K$ --
  \citet{HLL01}.}
\tablenotetext{e}{$y$ magnitude.}
\tablenotetext{f}{$B$ magnitude.}
\tablenotetext{g}{Out-of-eclipse maximum.}
\end{deluxetable}

\renewcommand{\arraystretch}{.6}
\begin{deluxetable}{llrllr}
\tablewidth{0pt}
\tablecaption{Properties of the Hottest Apparently Single White Dwarfs}
\tablehead{ 
\multicolumn{1}{l}{WD Number} & 
\multicolumn{1}{c}{Spectral Type\,\tablenotemark{a}} & 
\multicolumn{1}{c}{$T_{\rm eff}$\,\tablenotemark{b}} & 
\multicolumn{1}{c}{$V$\,\tablenotemark{a}} & 
\multicolumn{1}{c}{X-Ray\,\tablenotemark{c}} &
\multicolumn{1}{c}{Exp.\ Time\,\tablenotemark{d}} \\
\multicolumn{1}{c}{} &  
\multicolumn{1}{c}{} & 
\multicolumn{1}{c}{[K]} & 
\multicolumn{1}{c}{[Mag]} & 
\multicolumn{1}{c}{} & 
\multicolumn{1}{c}{[ks]}    
}
\startdata
WD\,0005$+$511  & ~~DO       & 120,000~~ & ~~13.32&  ~S, H  & 5.0~~ \\ 
WD\,0044$-$121  & ~~PG\,1159 & 150,000~~ & ~~11.84& ~S      & 11.4~~ \\
WD\,0122$-$753J & ~~DO       & 180,000~~ & ~~15.4 & ~S, H?   &  5.6~~ \\
WD\,0726+133    & ~~PG\,1159 & 140,000\,\tablenotemark{e}\,~ & ~~15.99& ~ND &  2.8~~ \\
WD\,1144$+$004  & ~~DQZO1    & 150,000~~ & ~~15.10& ~S      &  5.8~~ \\
WD\,1159$-$034  & ~~DQZO.4   & 140,000~~ & ~~14.84& ~S, H   & 13.6~~ \\
WD\,1501$+$664  & ~~DZ1      & 170,000~~ & ~~15.9 & ~S      & 43.2~~ \\
WD\,1522$+$662  & ~~DO       & 140,000~~ & ~~16.4\,\tablenotemark{f}   & ~S  & 4.7~~ \\
WD\,1821$+$643  & ~~DOZ.4    & 140,000\,\tablenotemark{g}\,~ & ~~15.04 & ~S  &  3.1~~ \\
WD\,1957$+$225  & ~~DAO      & 108,600~~ & ~~14.2 & ~S      & 19.9~~ \\
WD\,2117$+$342J & ~~DO       & 170,000~~ & ~~13.16& ~S      & 26.6~~ \\
WD\,2226$-$210  & ~~DAO      & 103,600~~ & ~~13.4 & ~S, H    &  4.9~~ \\
WD\,2333+301    & ~~DOZ.3    & 150,000\,\tablenotemark{e}\,~ & ~~16.13& ~ND &  4.2~~ \\
\enddata
\tablenotetext{a}{Visual magniture from \citet{MCC99}, unless noted otherwise.}
\tablenotetext{b}{Stellar effective temperature from \citet{Nap99}, unless noted otherwise.}
\tablenotetext{c}{X-ray emission detected by $ROSAT$ PSPC observations. ND: not detected;
  S: detected in the 0.1--0.5 keV band; H: detected in the 0.6--2.4 keV band.} 
\tablenotetext{d}{Exposure time of available $ROSAT$ PSPC observation.}
\tablenotetext{e}{From \citet{Wetal97}.}
\tablenotetext{f}{$B$ magnitude from \citet{MCC99}.}
\tablenotetext{g}{From \citet{KDR98}.}
\end{deluxetable}

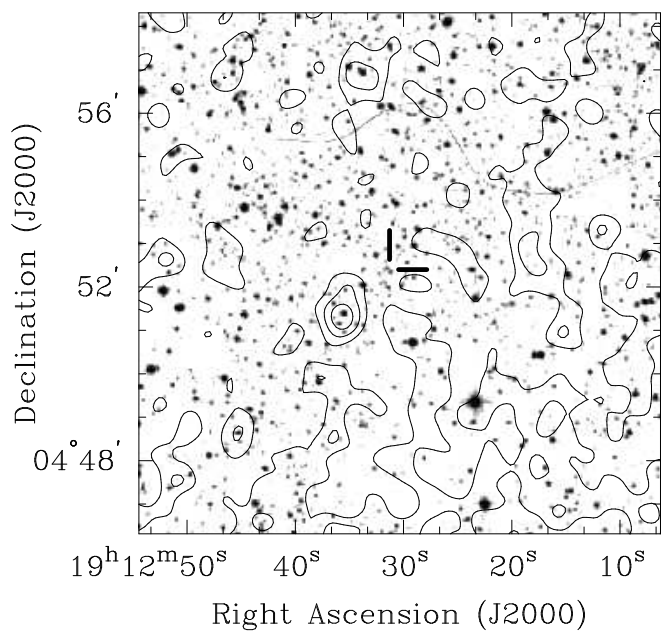
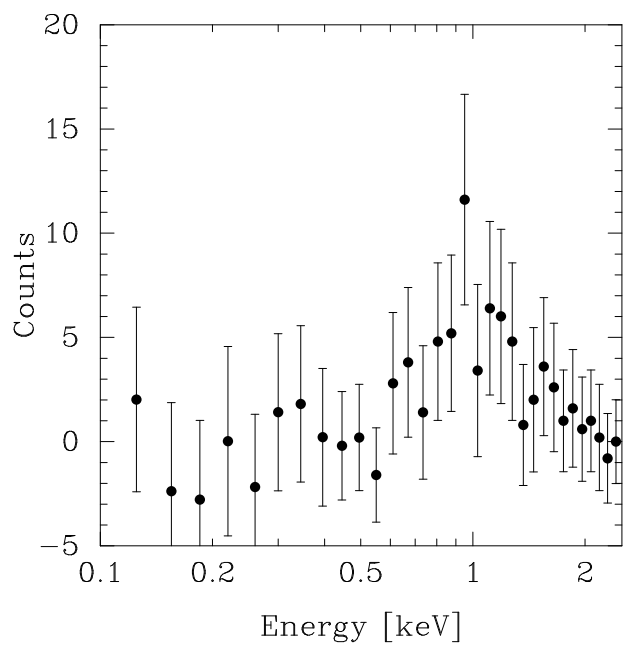
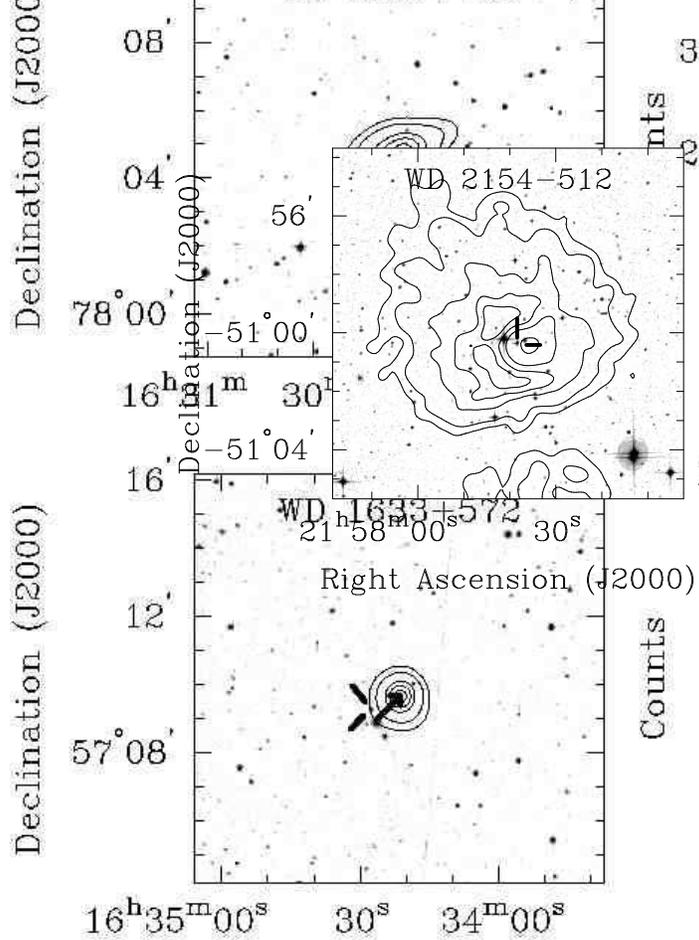
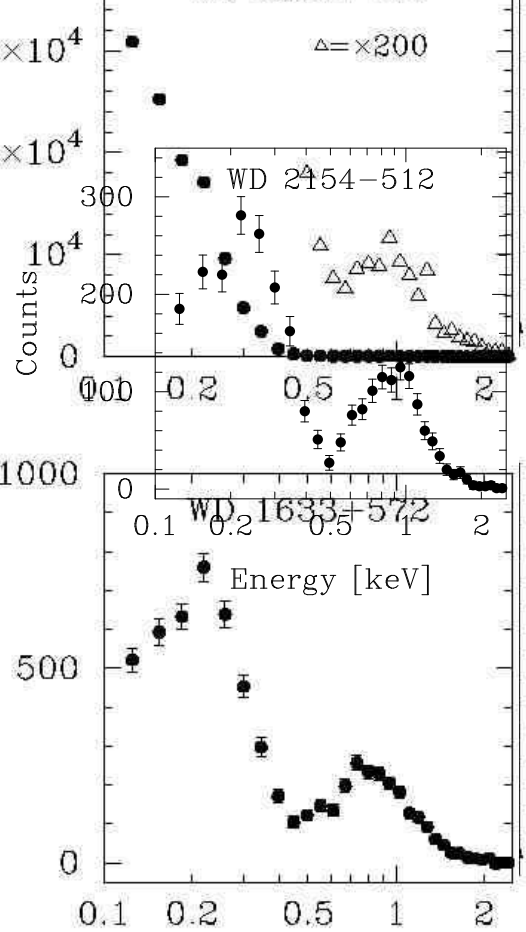
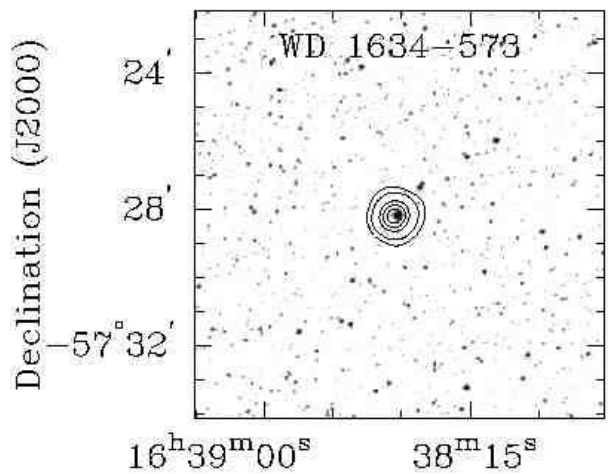
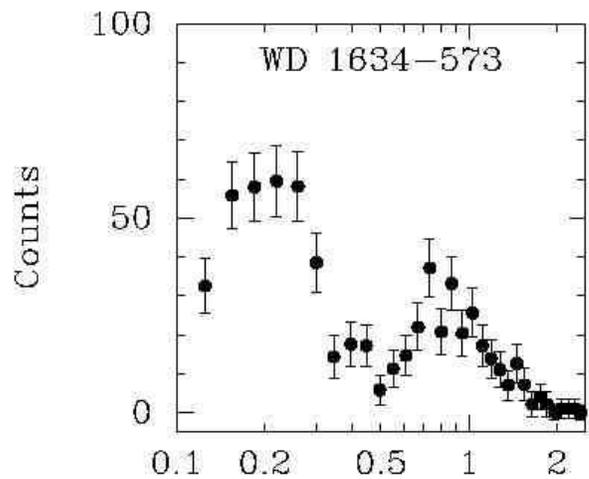
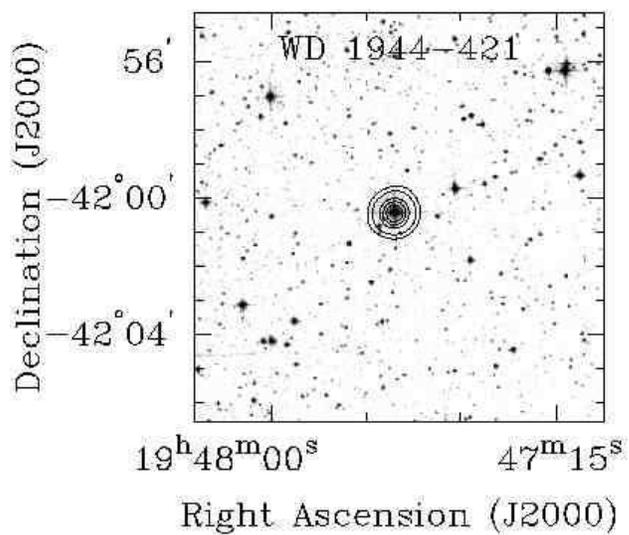
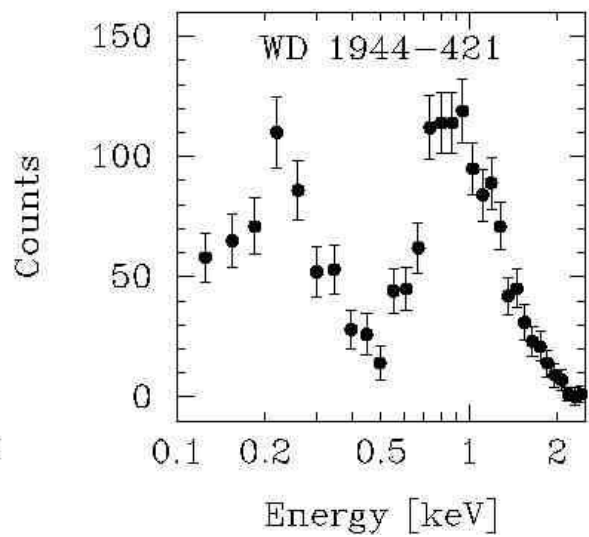
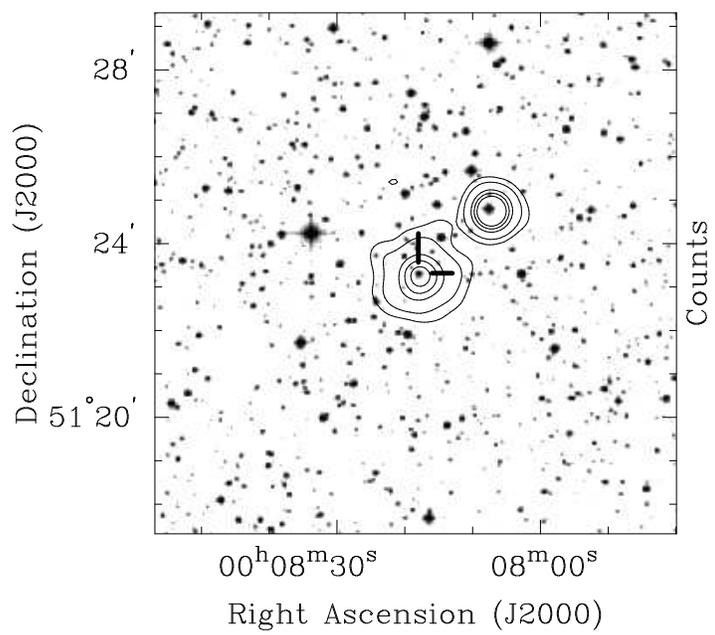
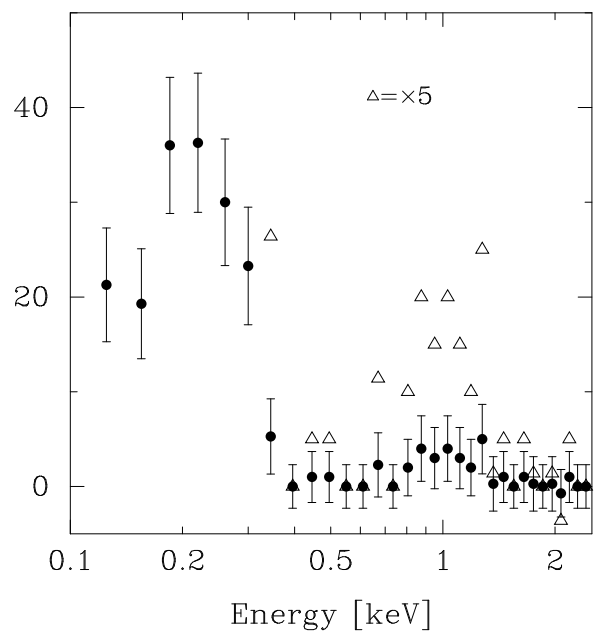
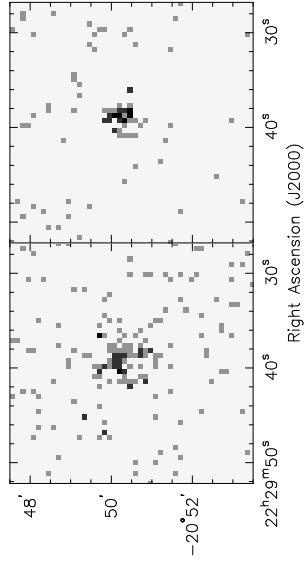
\end{document}